\documentclass[12pt,preprint]{aastex}

\slugcomment{To appear in ApJ Letters}

\shorttitle{Second-generation AGBs in GCs}
\shortauthors{Garc\'{\i}a-Hern\'andez et al.}

\begin{document}


\title{Clear evidence for the presence of second-generation asymptotic giant
branch stars in metal-poor Galactic globular clusters}


\author{D. A. Garc\'{\i}a-Hern\'andez\altaffilmark{1,2}, Sz.
M\'esz\'aros\altaffilmark{3}, M. Monelli\altaffilmark{1,2}, S.
Cassisi\altaffilmark{1,2,4}, P. B. Stetson\altaffilmark{5}, O.
Zamora\altaffilmark{1,2}, M. Shetrone\altaffilmark{6}, S.
Lucatello\altaffilmark{7}}


\altaffiltext{1}{Instituto de Astrof\'{\i}sica de Canarias, C/ Via L\'actea s/n, E-38205 La Laguna, Spain; agarcia@iac.es, monelli@iac.es, ozamora@iac.es}
\altaffiltext{2}{Departamento de Astrof\'{\i}sica, Universidad de La Laguna (ULL), E-38206 La Laguna, Spain}
\altaffiltext{3}{ELTE Gothard Astrophysical Observatory, H-9704 Szombathely, Szent Imre herceg \'ut, Hungary}
\altaffiltext{4}{INAF $-$ Osservatorio Astronomico di Teramo, via M. Maggini, 64100, Teramo, Italy; cassisi@oa-teramo.inaf.it}
\altaffiltext{5}{Dominion Astrophysical Observatory, Herzberg Institute of Astrophysics, National Research Council, 5071 West Saanich Road, Victoria, BC V9E 2E7, Canada}
\altaffiltext{6}{University of Texas at Austin, McDonald Observatory, Fort Davis, TX 79734, USA}
\altaffiltext{7}{INAF-Osservatorio Astronomico di Padova, vicolo dell Osservatorio 5, I-35122 Padova, Italy}


\begin{abstract}
Galactic globular clusters (GCs) are known to host multiple stellar populations:
a first generation with a chemical pattern typical of halo field stars and a
second generation (SG) enriched in Na and Al and depleted in O and Mg. Both
stellar generations are found at different evolutionary stages (e.g., the
main-sequence turnoff, the subgiant branch, and the red giant branch). The non
detection of SG asymptotic giant branch (AGB) stars in several metal-poor
([Fe/H] $<-1$) GCs suggests that not all SG stars  ascend the AGB phase,  and
that failed AGB stars may be very common in metal-poor GCs. This observation
represents a serious problem for stellar evolution and GC formation/evolution
theories. We report fourteen SG-AGB stars in four metal-poor GCs (M 13, M 5, M
3, and M 2) with different observational properties: horizontal branch (HB)
morphology, metallicity, and age. By combining the H-band Al abundances obtained
by the  APOGEE survey with ground-based optical photometry, we identify SG
Al-rich AGB stars in these four GCs and show that Al-rich RGB/AGB GC stars
should be Na-rich. Our observations provide strong support for present,
standard stellar models, i.e., without including a strong mass-loss efficiency,
for low-mass HB stars. In fact, current empirical evidence is in agreement with
the predicted distribution of FG and and SG stars during the He-burning stages
based on these standard stellar models.
\end{abstract}





\section{Introduction}

It is well known that all Galactic globular clusters (GCs) host multiple (at
least two) stellar populations (see Gratton et al. 2012; Piotto et al. 2012, and
references therein). This result has been deduced mainly from the general presence of
the so-called C$-$N, O$-$Na, and Mg$-$Al anticorrelations. First-generation (FG)
stars display normal Na (and Al) abundances (i.e., typical of halo field
stars), while second-generation (SG) stars - which may have additional
subpopulations - show Na (and Al) enhancements. These SG additional
subpopulations of stars are also characterized by He overabundances that change
from cluster to cluster (e.g., Milone et al. 2014). The presence of first- and
second-generation stars in GCs has been clearly traced - using both
spectroscopic and/or photometric data - in the various evolutionary
sequences, from the main sequence up to the more advanced evolutionary stages
(e.g., Gratton et al. 2001; Carretta et al. 2009; Milone et al. 2012; Marino et
al. 2014). However, as initially noticed by Norris et al. (1981, see also
Gratton et al. 2010), there are claims concerning the paucity (or lack) of SG
stars along the asymptotic giant branch (AGB) of some clusters. If confirmed,
this occurrence would represent a challenge for stellar evolution and the
formation and evolution models of these complex stellar systems (Charbonnel et
al. 2013; Cassisi et al. 2014).

Recent spectroscopic observations of AGB stars in the metal-poor
([Fe/H] $\approx -1.56$) GC NGC 6752 (Campbell et al. 2013) have found no Na-rich
SG-AGB stars in this cluster.\footnote{See also Johnson \& Pilachowski (2012) and
Lapenna et al.(2015) for the non-detection of SG-AGBs in M 13 (a twin of NGC
6752) and M 62 (a slightly more metal-rich cluster with [Fe/H] $\approx-1.2$).}
Campbell et al. (2013) have explained these puzzling observations as due to the fact
that all SG stars do not ascend the AGB (AGB-manqu\'e stars). They suggest that
a stronger mass-loss in SG horizontal branch (HB) stars could explain their
observations, and that AGB-manqu\'e stars may be very common in metal-poor
([Fe/H] $<-1$) Galactic GCs.\footnote{Johnson et al. (2015) have very recently
found SG (Na-rich) AGB stars in the massive and more metal-rich
([Fe/H ] $\approx-0.70$) GC 47 Tuc but they argue that the high metallicity leads
to a different HB morphology (predominantly red) with an insignificant
population of hot HB stars and failed AGB stars, which likely make up the
missing Na-rich AGB in NGC 6752 (and M 13).}

The Na-poor nature of all AGB stars analyzed in NGC 6752 (also in M 13 and M 62)
poses an apparent problem for stellar evolution. This difficulty arises
because synthetic HB models of NGC 6752, based on canonical - i.e., without a
strong mass-loss efficiency during the core He-burning stage -  HB stellar
models, do not predict the observed lack of SG-AGB stars. More recently,
Cassisi et al. (2014) have critically discussed such a mass-loss scenario during
the core He-burning stage. They show that the required mass-loss rates are much
higher than any of the current theoretical and empirical constraints, and that
if all SG HB stars do not climb the AGB it would be virtually impossible to
reproduce the number ratio of AGB to HB stars (the R$_{2}$ parameter) in NGC
6752 and a few other clusters with similar/dissimilar observational properties.
Thus, at present there is no simple explanation for the apparent lack of SG-AGB
stars in these metal-poor GCs.

Here we report SG-AGB stars in the GC M 13 - a twin of NGC 6752 - and another
three GCs (M 5, M 3, and M 2) of similar metallicity but with distinct
observational properties in terms of HB morphology and age. For this, we
combined the H-band  abundances measured by the Apache Point Observatory
Galactic Evolution Experiment (APOGEE; Majewski et al. 2015) and the most recent
ground-based photometry of these GCs. The C and N abundances are
significantly affected by the occurrence of both the first dredge-up during the
early RGB and by non-canonical extra-mixing processes after the RGB bump (e.g.,
Cassisi \& Salaris 2013). Thus, for the present analysis we decided to analyze
the abundances of Mg, Al, Na, and O which are barely - if at all - affected by
these mixing processes during the RGB and AGB stages in the evolution of
low-mass stars.

\section{APOGEE data and ground-based photometry}

The APOGEE survey observed ten northern GCs, covering a range of metallicity [Fe/H] from
$-0.8$ down to $-2.4$, including cluster members with well-characterized stellar
parameters and abundances from existing high-resolution optical spectra, as well
as many additional cluster giant stars currently lacking such detailed
abundances (M\'esz\'aros et al. 2015). The stellar parameters and chemical
abundances of nine elements (Fe, C, N, O, Mg, Al, Si, Ca, and Ti) for 428
cluster star members in these ten GCs have recently been reported by us
(M\'esz\'aros et al. 2015). We used photometry and theoretical isochrones to
constrain the effective temperature ($T_{\rm eff}$) and surface gravity ($\log g$). We
then used an independent semi-automated method for precise (up to the $\sim$0.1
dex level) elemental abundance determination from the high-resolution
($\sim$22\,500) and high-quality (signal-to-noise $>$ 70 per pixel) $H$-band
($\sim$1.5--1.7 $\mu$m) spectra (M\'esz\'aros et al. 2015).   

The APOGEE abundances are measured from neutral lines of Fe, Al, Mg, etc.; the
$H$-band single-ionized lines are not detected in metal-poor GC giants. The APOGEE
$H$-band data offer several advantages with respect to previous optical
spectroscopic studies of GC giants: i) they enable us to analyze these ten clusters
in a homogeneous way (covering almost the full extent of the RGB); ii) nonlocal
thermodynamic equilibrium (NLTE) effects on the spectral lines of neutral
species such as Fe, Al, Mg, etc. are less important than in the optical range
because in the $H$-band these lines are formed deeper in the atmosphere (see Sect.
4); and iii) the increased number of APOGEE stars compared to the literature
permit us to discover more Al-rich stars, making the Mg--Al plane clearer in
the APOGEE data than in previous studies.

Ground-based $U,B,V,I$ photometry is available for six of the ten GCs observed
by APOGEE. The ground-based photometry is taken from the private collection by
P. Stetson, which is based upon a large corpus of the most recent observations
obtained mainly from public astronomical archives. Our $U$ and $BVI$ magnitudes are
precise to the level of $<$0.002 and $<$0.001 mag, respectively, for stars in
the range ($11.5\leq V\leq 15.5$) of the APOGEE observations (see e.g., Stetson
et al. 2014).

\section{Color--magnitude diagrams and the ($V$, $C_{u,b,i}$) pseudo-CMD}

We made use of the ground-based $U,B,V,I$ photometry mentioned above to construct
several color--magnitude diagrams (CMDs) for each GC observed by APOGEE and
separate the AGB from the RGB stars.\footnote{There is no significant
differential reddening in the APOGEE GCs with available ground-based optical
photometry, and the separation of AGB stars from RGB stars is easier.} We find
that the combination of the $U$--$(U-I)$, I$-$(U-I), and $V$--$(B-I)$ CMDs gives an
efficient RGB/AGB separation (Figure 1; see below). 

In M\'esz\'aros et al. (2015) we used an extreme-deconvolution (XD)
method\footnote{http://github.com/jobovy/extreme-deconvolution} to identify FG
and SG stars in the Al--Mg distributions and to assign cluster membership.
Briefly, the XD method fits the distribution of the elemental abundances as a
sum of K Gaussian populations. Similar to K-means (Steinhaus 1956), the number
of populations to fit is an input to the XD method and was fixed to two
populations (FG and SG). This XD method was applied to each GC in our sample by
using [Mg/Fe], [Al/Fe], [Si/Fe], [Ca/Fe], and [Ti/Fe], as well as only [Mg/Fe]
and [Al/Fe] (taking into account the individual abundance errors), and we found
that Mg and Al drive the population membership in the most metal-poor ([Fe/H]
$<-1$) GCs (see M\'esz\'aros et al. 2015 for more details). The exact [Al/Fe]
boundary between FG and SG stars is thus provided by the XD method and may
differ slightly from one cluster to another. We note that (in the four GCs with
SG-AGB stars identified, see below), this basically translates into FG and SG
stars displaying roughly [Al/Fe] $<$ 0.50 dex (Al-poor) and [Al/Fe] $\ge$ 0.50
dex (Al-rich), respectively. Thus, we combined our FG and SG star classification
(mainly driven by the Al abundances) with the $U$--$(U-I)$, $I$--$(U-I)$, and
$V$--$(B-I)$ CMDs. We note that FG- and SG-AGB stars display [Al/Fe] $<$ 0.50
dex (Al-poor) and [Al/Fe] $\ge$ 0.50 dex (Al-rich), respectively. The only
exception is the M 2 AGB star 2M21331521-0049516, which displays a slightly
lower Al abundance ([Al/Fe] = 0.37 dex) and is classified as a SG star by the XD
method. 

The CMDs for four GCs (M 13: [Fe/H]$\approx$-1.53, 67 stars; M 5:
[Fe/H]$\approx$-1.29, 103 stars; M 3: [Fe/H]$\approx$-1.50, 46 stars; and M 2:
[Fe/H]$\approx$-1.65, 18 stars) contain SG Al-rich AGB stars (Figure 1). The
AGB stars are clearly separated from those of the RGB in the $U$--$(U-I)$, $I$--$(U-I)$,
and $V$--$(B-I)$ CMDs (Figure 1). The only exceptions are: i) four M 13 AGB stars
(the brightest ones near the tip of the RGB; all of them FG), which are not
clearly separated from the RGB stars in the $V$--$(B-I)$ CMD but are in the
$I$--$(U-I)$ and $U$--$(U-I)$ CMDs; ii) two M 3 AGB stars (both SG) that lie on the
red RGB tail in the $U$--$(U-I)$ CMD but are clearly identified as AGB stars
in the other two CMDs\footnote{Our RGB/AGB identification is more conservative
in M 3 because the RGB is not so well defined as in the case of M 13 and M 5.};
and iii) one M 2 AGB star (an FG one) that is not well separated from the RGBs in
the $I$--$(U-I)$ and $V$--$(B-I)$ CMDs but is in the $U$--$(U-I)$ CMD. We identify a
total of 4, 5, 3, and 2 SG Al-rich AGB stars in M 13, M 5, M 3, and M 2,
respectively. Table 1 lists the AGB stars (both FG and SG) identified together
with some relevant observational information such as the APOGEE Al (and O where
available) abundances and the Na and O abundances from the
literature.\footnote{There is only one Na I line (1.639 $\mu$m) in the APOGEE
$H$-band spectral range, which is too weak in the spectra of low-metallicity GC
giant stars for reliable abundances to be derived.} 

The Al--O anticorrelation is clearly seen in our APOGEE data for all clusters
(Figure 2) and the SG Al-rich AGB stars are among the most O-poor
stars\footnote{We use the most recent Asplund et al. (2005) solar abundance
scale (e.g., A(O)=8.66), while earlier literature optical works generally use
older solar abundance scales (e.g., Anders \& Grevesse 1989; Grevesse \& Sauval
1998). The mean offset of $\sim$ $+$0.2--0.3 dex between the APOGEE and
literature O abundances is just the consequence of using different solar
abundance scales (M{\'e}sz{\'a}ros et al. 2015). Using older solar abundance
scales, our APOGEE O abundances would be in good agreement with the literature; in
particular the [O/Fe] abundances in SG Al-rich AGB stars would be $\leq$0.0.},
as expected. Only a few (5 out of 14) of the SG Al-rich AGB stars have Na
abundances from optical spectroscopy available in the literature (Table 1).
Remarkably, all of them are Na-rich ([Na/Fe] $\sim$ 0.3--0.6 dex; see Figure 3),
supporting their identification as truly SG-AGB stars. Another indication of the
Na-rich nature of the identified SG-AGB stars is offered by the ($V$, $C_{u,b,i}$)
pseudo-CMDs (where $C_{u,b,i}=(U-B)-(B-I)$; Monelli et al. 2013). It has been
clearly shown by Monelli et al. (2013) that the $C_{u,b,i}$ index is very
sensitive to any change in the relative distributions of CNO elements, and, since
SG stars are N-rich/O-poor/Na-rich/Al-rich with respect FG stars, it is a
powerful tool for tracing the distribution of FG/SG stars along the RGB and AGB
evolutionary stages. In order to find an independent confirmation of present
results, we show in Figure 4 the ($V$, $C_{u,b,i}$) pseudo-CMD of all the GCs in our
sample. We find that (on average, with some exceptions) both RGB and AGB stars
in our GC sample are separated in the ($V$, $C_{u,b,i}$) pseudo-CMDs depending on
their Al content (FG or SG). The SG Al-rich stars generally display higher
values of the $C_{u,b,i}$ index, which corresponds to a population with
higher Na content (Monelli et al. 2013). For example, at least two SG-AGB stars
in M 13 (2M16412975$+$3631563 and 2M16414398$+$3622338) lie in the region
occupied by the most extreme Na-rich population defined by Monelli et al.
(2013). 

Finally, another interesting feature of Figure 1 is a hint for the presence
of a splitting (i.e., different photometric sequences) along the AGB between the
FG- and SG-AGB stars in M 13 and M 5. The number of stars, however, is
small and this AGB splitting is not seen in M 3 and M 2, where we have 
observed even fewer stars. In the CMDs, the M 13 and M 5 SG-AGB stars seem to define bluer
(and/or brighter) photometric sequences than the FG ones, as expected. 

\section{Discussion and conclusions}

The non-detection of SG-AGB stars in several metal-poor GCs (such as NGC 6752, M
13, and M 62) from previous optical spectroscopic surveys may be just
coincidental (bias in the sample selections, small stellar samples) or due to
the non-use of recent (and precise) optical photometry and appropriate
combinations of several CMDs for efficient RGB/AGB separation. For example, we
have several M 13 AGB stars (9 FG and 1 SG; see Table 1) in common with Johnson
\& Pilachowski (2012). These authors used the $V$--$(V-K)$ CMD (with coordinates
and $V$ photometric data by Cudworth \& Monet 1979) to separate the RGB from
the AGB stars. They are not able to efficiently discriminate the RGB from the AGB
(especially near the tip of the RGB) because the width of the RGB in their
$V$--$(V-K)$ CMD is much wider than ours (when using our more recent photometric
data). Indeed, the only M 13 SG-AGB star (2M16414398$+$3622338; Table 1) in
common with us was wrongly classified by these authors as an RGB star. Our AGB
identifications in M 5 are fully consistent with the previous optical studies
using recent photometric data. The previous optical works in M 3 did not attempt
any RGB/AGB separation from appropriate combinations of several CMDs, while no M
2 AGB star in our sample has been previously studied (see Table 1).

The lack of Na-rich SG-AGB stars in NGC 6752 (Campbell et al. 2013) is puzzling
(also in M 62 but only six AGB stars were analyzed; Lapenna et al. 2015). Here
we report for the first time SG-AGB stars in the GC M 13; a twin of NGC 6752
with very similar HB morphology, metallicity, and age. An alternative
explanation for the Na-poor character of all AGBs surveyed in NGC 6752 (as well
as for the previous non-detection of Na-rich SG-AGBs in several metal-poor GCs)
is the fact that NLTE effects in AGB stars may be larger than in RGB stars. This
would underestimate more severely the correct Na abundances in the AGB stars
(Lapenna et al. 2015 and references therein). Higher NLTE effects in AGB stars
are suggested by the differences (up to $\sim$0.1--0.2 dex) in the Fe (and Ti)
abundances measured from neutral and single-ionized lines in the AGB stars,
which otherwise are negligible in the RGB stars (e.g., Ivans et al. 2001;
Lapenna et al. 2015). For example, the Fe abundances derived from optical
neutral lines in AGB stars are systematically $\sim$0.1--0.2 dex lower than in
the RGB stars. As we mentioned above, the APOGEE abundances are measured from
neutral lines and we find no significant differences for the Fe (Al, Mg, O)
abundances in the AGB and RGB stars (see Figure 2). This confirms that the
$H$-band spectral lines are formed deeper in the atmosphere and NLTE effects on
the neutral lines of Fe, Al, Mg, etc. are less severe than in the optical
domain. The $H$-band thus opens up a new (and safer) window to systematically study
the AGB and RGB stellar generations in Galactic GCs.

In conclusion, our results provide plain evidence of the fact that SG
stars are present along the AGB of metal-poor Galactic GCs. This supports the
present generation of canonical HB stellar models in terms of their capability
to properly reproduce the observed distribution of FG and SG stars during both
the core and shell He-burning phases.



\acknowledgments

DAGH/OZ acknowledge MINECO support under grant AYA$-$2014$-$58082-P. SzM is
supported by the J{\'a}nos Bolyai Research Scholarship of the Hungarian Academy
of Sciences. SC acknowledges partial financial support from PRIN-INAF2014 and the IAC for inviting him
as a Severo Ochoa visitor during April to June 2015 when part of this work was
done. SL acknowledges partial support from PRIN-MIUR 2010-2011. Funding for SDSS-III has been provided by the Alfred P. Sloan
Foundation, the Participating Institutions, the National Science Foundation, and
the U.S. Department of Energy Office of Science. The SDSS-III web site is
http://www.sdss3.org/. SDSS-III is managed by the Astrophysical Research
Consortium for the Participating Institutions of the SDSS-III Collaboration
including the University of Arizona, the Brazilian Participation Group,
Brookhaven National Laboratory, University of Cambridge, Carnegie Mellon
University, University of Florida, the French Participation Group, the German
Participation Group, Harvard University, the Instituto de Astrof{\'{\i}}sica de
Canarias, the Michigan State/Notre Dame/JINA Participation Group, Johns Hopkins
University, Lawrence Berkeley National Laboratory, Max Planck Institute for
Astrophysics, New Mexico State University, New York University, Ohio State
University, Pennsylvania State University, University of Portsmouth, Princeton
University, the Spanish Participation Group, University of Tokyo, University of
Utah, Vanderbilt University, University of Virginia, University of Washington,
and Yale University.



{\it Facilities:} \facility{SDSS-III:APOGEE}.

\clearpage

\begin{deluxetable}{lcccccccc}
\tabletypesize{\scriptsize}
\tablecaption{AGB stars in metal-poor globular clusters$^{a}$ \label{tbl-1}}
\tablewidth{0pt}
\tablehead{
\colhead{2MASS Name} &   \colhead{T$_{eff}$} & \colhead{log g} &\colhead{Pop.$^{b}$} & \colhead{[Al/Fe]} &
\colhead{[O/Fe]$^{c}$} & \colhead{[Na/Fe]$_{Lit.}$$^{d}$} & \colhead{[O/Fe]$_{Lit.}$$^{d}$}& \colhead{Ref$^{e}$.} \\
}
\startdata
  &  &  & M 13 &  &  &   \\
\hline
2M16422126+3633533 & 5136 & 2.59 & 1 & -0.39 & $\dots$  & 0.20  (0.26)  & 0.69     & 1  \\
2M16412975+3631563 & 5173 & 2.67 & 2 &  0.82 & $\dots$  & $\dots$       & $\dots$  & $\dots$  \\
2M16415003+3625105 & 4698 & 1.65 & 1 &  0.26 & $\dots$  & 0.30  (0.18)  & 0.25     & 1 \\
2M16415543+3633266 & 5024 & 2.33 & 2 &  0.75 & $\dots$  & $\dots$       & $\dots$  & $\dots$  \\
2M16415024+3629431 & 4909 & 2.08 & 2 &  0.59 & $\dots$  & $\dots$       & $\dots$  & $\dots$  \\
2M16415452+3626289 & 5376 & 3.23 & 1 &  0.42 & $\dots$  & $\dots$       & $\dots$  & $\dots$  \\
2M16413082+3630130 & 4950 & 2.16 & 1 &  0.04 & $\dots$  & 0.05          & 0.29     & 6  \\
2M16414398+3622338 & 4606 & 1.51 & 2 &  0.72 & $\dots$  & 0.29  (0.13)  & 0.10     & 1  \\
2M16420085+3623338 & 4594 & 1.45 & 1 &  0.16 & $\dots$  & 0.22  (0.06)  & 0.32     & 1  \\
2M16412408+3625306 & 4366 & 1.08 & 1 &  0.20 & 0.60     & 0.00  (-0.16) & 0.14     & 1  \\
2M16412709+3628002 & 4366 & 1.08 & 1 & -0.25 & 0.59     & -0.09         & 0.30     & 6  \\
2M16413961+3627381 & 4337 & 1.03 & 1 & -0.14 & 0.58     & 0.01  (-0.15) & 0.38     & 1  \\
2M16414966+3627104 & 4512 & 1.32 & 1 & -0.30 & 0.60     & -0.26 (-0.42) & 0.46     & 1  \\
2M16414517+3628132 & 4435 & 1.17 & 1 & -0.15 & 0.54     & 0.25  (0.09)  & 0.34     & 1  \\
\hline
  &  &  & M 5 &  &  &   \\
\hline
2M15184048+0210446  & 5499 & 3.58 & 1 & -0.28 & $\dots$  &$\dots$     & $\dots$ & $\dots$ \\
2M15180831+0158530  & 4922 & 2.30 & 1 &  0.10 & $\dots$  &0.35        & 0.30    & 2 \\
2M15183957+0205018  & 5071 & 2.63 & 2 &  0.50 & $\dots$  &$\dots$     & $\dots$ & $\dots$ \\
2M15184022+0213278  & 4966 & 2.41 & 1 & -0.26 & $\dots$  &$\dots$     & $\dots$ & $\dots$ \\
2M15175224+0208026  & 5078 & 2.63 & 1 &  0.24 & $\dots$  &$\dots$     & $\dots$ & $\dots$ \\
2M15185731+0203077  & 5067 & 2.63 & 1 &  0.26 & $\dots$  &$\dots$     & $\dots$ & $\dots$ \\
2M15183638+0208507  & 4842 & 2.12 & 2 &  0.84 & $\dots$  &$\dots$     & $\dots$ & $\dots$ \\
2M15180987+0210088  & 4810 & 2.06 & 1 & -0.05 & $\dots$  &-0.04       & 0.56    & 2       \\
2M15183575+0204297  & 6155 & 3.87 & 2 &  1.06 & $\dots$  &$\dots$     & $\dots$ & $\dots$ \\
2M15185515+0214337  & 4639 & 1.73 & 1 & -0.15 & $\dots$  &-0.01       & 0.31    & 2       \\
2M15182435+0201574  & 4507 & 1.53 & 2 &  0.90 & 0.03     &0.41 (0.25) & -0.11   & 3      \\
2M15183738+0206079  & 4207 & 0.96 & 2 &  0.50 & 0.29     &0.29        & 0.28    & 2       \\
2M15174702+0204519  & 4391 & 1.30 & 1 &  -0.10 & 0.44 & -0.08         & 0.47 & 2 \\
2M15182014+0203321  & 4533 & 1.53 & 1 &   0.09 & 0.32 &  0.20 (0.04)  & 0.13 & 3 \\
2M15184540+0204302  & 4283 & 1.10 & 1 &   0.15 & 0.41 &  0.26         & 0.49 & 2 \\
2M15184139+0206004  & 4306 & 1.15 & 1 &   0.27 & 0.51 &  0.52 (0.33)  & 0.35 & 3 \\
\hline
  &  &  & M 3 &  &  &   \\
\hline
2M13423482+2826148 & 4904 & 2.13 & 1 & -0.24 & $\dots$  & -0.19        & $\dots$  & 4  \\
2M13414871+2820024 & 5011 & 2.36 & 2 &  0.77 & $\dots$  & $\dots$      & $\dots$  & $\dots$ \\
2M13421373+2821154 & 5032 & 2.40 & 1 & -0.19 & $\dots$  & 0.08         & $\dots$  & 4  \\
2M13422197+2828408 & 4700 & 1.71 & 1 & -0.12 & $\dots$  & -0.19        & $\dots$  & 4  \\
2M13425083+2827576 & 4880 & 2.08 & 1 & -0.13 & $\dots$  & -0.17        & $\dots$  & 4  \\
2M13421712+2822137 & 4418 & 1.22 & 1 & -0.08 & 0.58     &  0.05        & $\dots$  & 4  \\
2M13414576+2824597 & 4315 & 1.00 & 1 & -0.15 & 0.67     & -0.26        & 0.36     & 4,5  \\
2M13415152+2823224 & 4015 & 0.50 & 2 &  0.78 & 0.12     &  0.57        & -0.24    & 6  \\
2M13421086+2823465 & 4047 & 0.56 & 2 &  0.78 & 0.23     &  0.61 (0.42) & -0.05    & 5  \\
\hline
  &  &  & M 2 &  &  &   \\
\hline
2M21332545-0047056 & 4710 & 1.63 & 1 & -0.25 & $\dots$  & $\dots$              & $\dots$          & $\dots$    \\  
2M21332531-0052511 & 4531 & 1.31 & 1 & -0.38 & 0.65     & $\dots$              & $\dots$          & $\dots$    \\ 
2M21331521-0049516 & 4554 & 1.35 & 2 &  0.37 & $\dots$  & $\dots$              & $\dots$          & $\dots$    \\ 
2M21333432-0051285 & 4436 & 1.14 & 1 & -0.34 & 0.63     & $\dots$              & $\dots$          & $\dots$    \\ 
2M21332527-0049386 & 4098 & 0.53 & 2 &  0.68 &-0.02     & $\dots$              & $\dots$          & $\dots$    \\
\enddata
\tablenotetext{a}{Effective temperatures ($T_{\rm eff}$), surface gravities (log $g$), Al, and O abundances from M{\'e}sz{\'a}ros et al.(2015).}
\tablenotetext{b}{Population: 1 and 2 are first-generation and second-generation, respectively.}
\tablenotetext{c}{[O/Fe] abundances from M{\'e}sz{\'a}ros et al.(2015); only
available for stars with $T_{\rm eff}$ below 4520 K. Note that our APOGEE [O/Fe] abundances
are $\sim$ $+$0.2--0.3 dex sistematically higher than the literature values
because of the use of different solar abundance scales (M{\'e}sz{\'a}ros et al.
2015).}
\tablenotetext{d}{Na and O abundances in the literature (from high-resolution
optical spectra). The Na abundances corrected for NLTE effects according to
Gratton et al. (1999) are given. Johnson \& Pilachowski (2012) and Ivans et al.
(2001) did not report the measured Na I 6154\AA~equivalent widths (EWs) for the M
13 and M 5 AGB stars, respectively, while Cavallo \& Nagar (2000) report an EW
of $\sim$34 m\AA~for one M 3 AGB star in our sample. Thus, for these stars we
assumed an average EW of 30 m\AA~(e.g., the EWs range in M 13 giant stars is
$\sim$10$-$50 m\AA; Cohen \& Melendez 2005) and [Fe/H]=-1.50 to make
conservative NLTE corrections and we list also the non-corrected Na abundances
in parenthesis.}
\tablenotetext{e}{Reference for the O and Na abundances from high-resolution optical spectra.}
\tablerefs{(1) Johnson \& Pilachowski (2012); (2) Lai et al. (2011); (3) Ivans et al. (2001); (4) Johnson et
al. (2005); (5) Cavallo \& Nagar (2000); (6) Sneden et al. (2004).}
\end{deluxetable}

\clearpage

\begin{figure}
\centering
\includegraphics[angle=0,scale=.75]{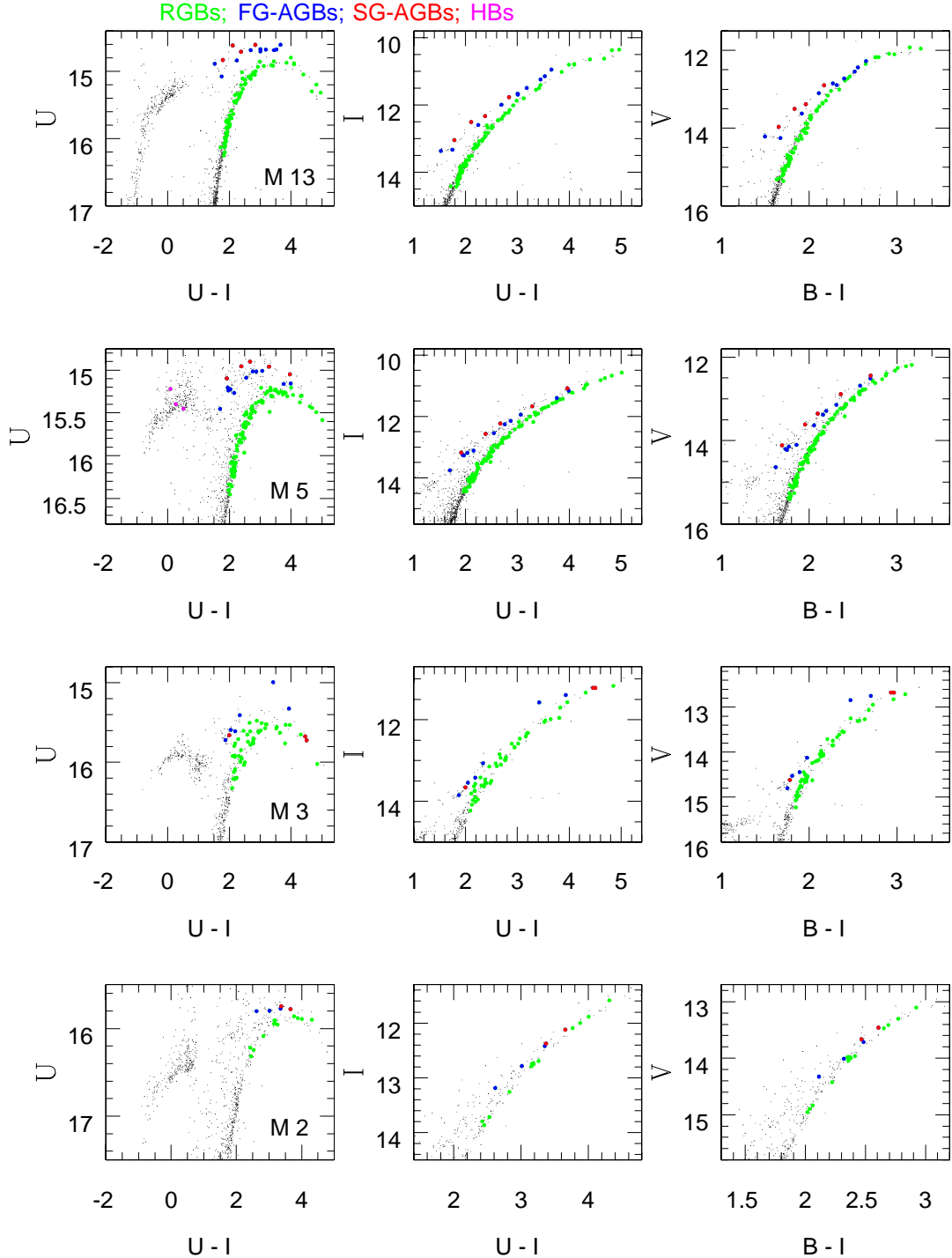}
\caption{Color--magnitude (CMD) diagrams $U$ vs. $(U-I)$ (left panels), $I$ vs. $(U-I)$
(middle panels), and $V$ vs. $(B-I)$ (right panels) for metal-poor GCs (from top to
bottom: M 13, M 5, M 3, and M 2). Ground-based photometry for the cluster stars
is indicated with black dots, while the RGB, FG-AGB, and SG-AGB stars observed
by APOGEE are indicated with green, blue, and red dots, respectively. The three
M 5 stars marked with magenta dots (left panel) are HB stars. \label{fig1}}
\end{figure}

\clearpage

\begin{figure}
\centering
\includegraphics[angle=-90,scale=.7]{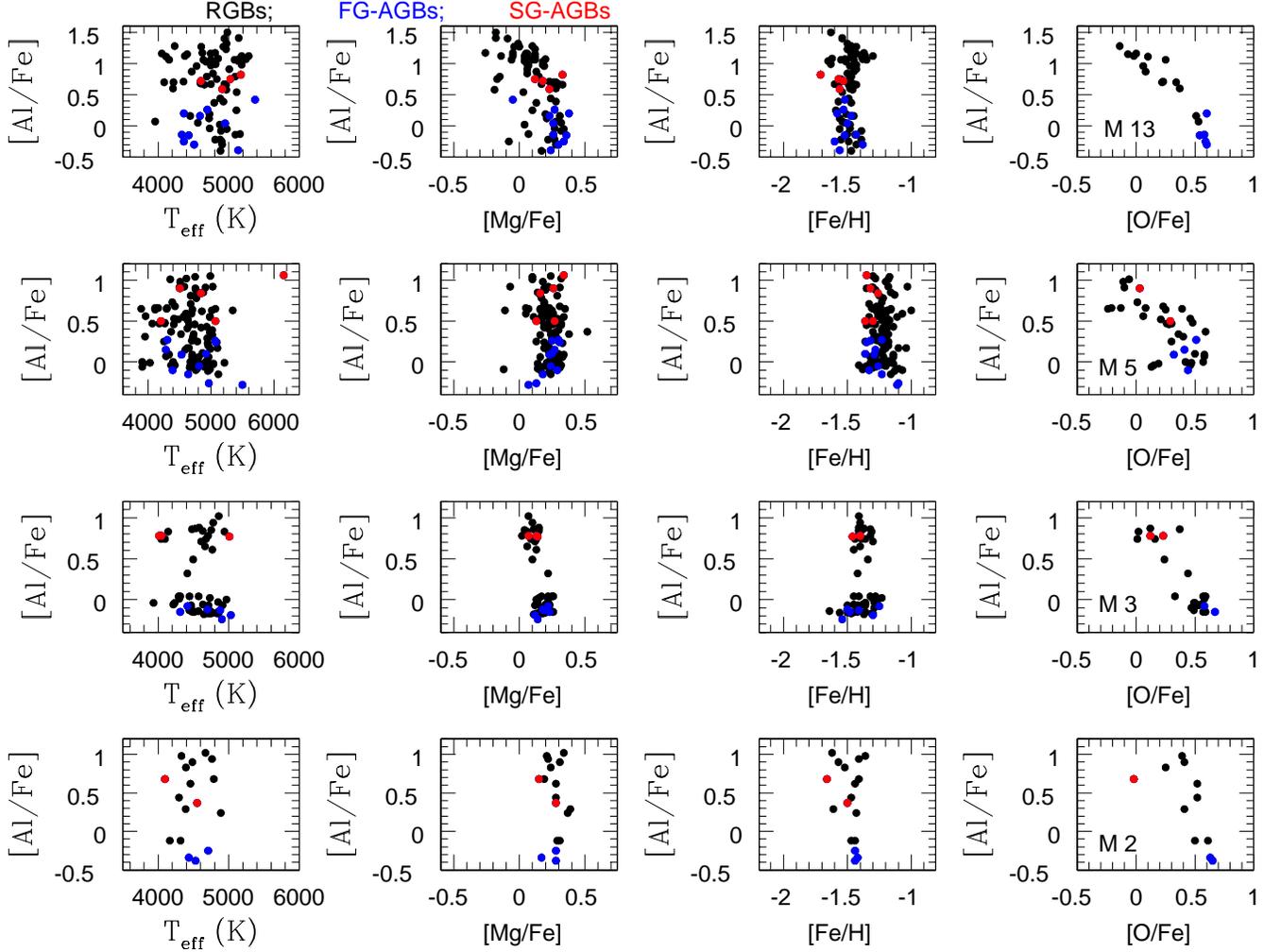}
\caption{[Al/Fe] for our APOGEE sample of FG- and SG-AGB stars (blue and red
dots, respectively) in metal-poor GC stars (from top to bottom: M 13, M 5, M3,
and M 2) shown against (from left to right) stellar effective temperature
$T_{\rm eff}$, [Mg/Fe], [Fe/H], and [O/Fe]. The APOGEE abundances are precise to the
$\sim$0.1 dex level. For comparison, the RGB stars (black dots) observed by
APOGEE are also displayed. Note that the spread in [Fe/H] is bigger in RGB than
in AGB stars because of the presence of warm stars, which lead to bigger
uncertainties (consequentely [Al/Fe] is determined with higher accuracy in AGB
than in RGB stars).  \label{fig2}}
\end{figure}
\clearpage

\begin{figure}
\centering
\includegraphics[angle=-90,scale=.5]{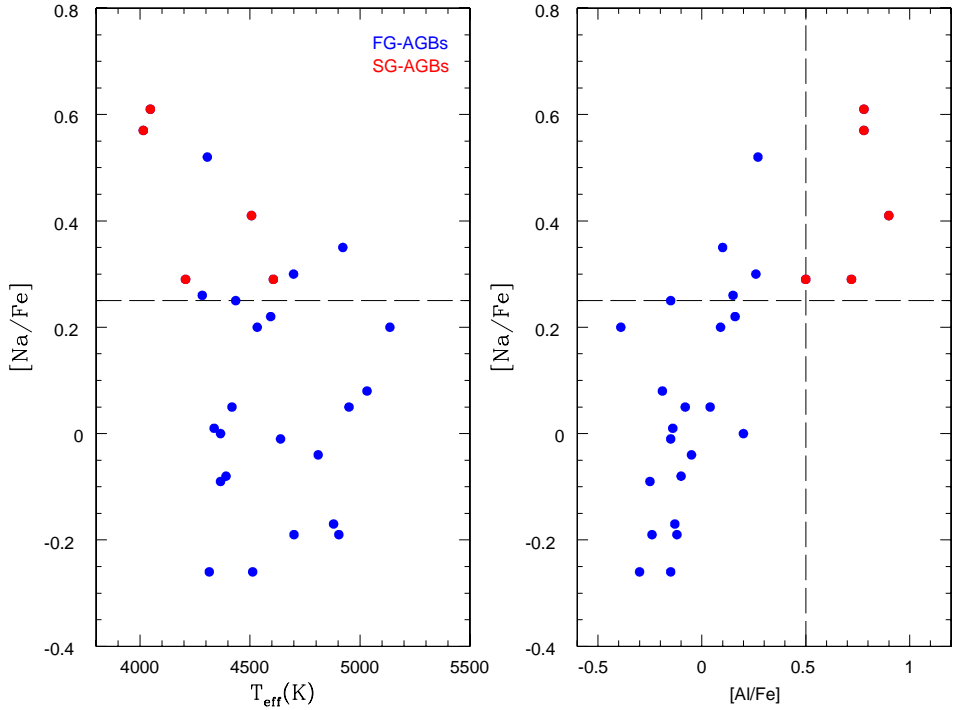}
\caption{Literature [Na/Fe] abundances (corrected for NLTE) for our APOGEE
sample of FG- and SG-AGB stars (blue and red dots, respectively) shown against
stellar effective temperature $T_{\rm eff}$ (left panel) and Al abundances (right
panel). The horizontal and vertical lines mark [Na/Fe]=$+$0.25 dex and
[Al/Fe] = $+$0.50 dex, respectively and separate the FG-AGBs from the SG ones.
The Na limit is set to [Na/Fe] = 0.25 dex, corresponding to the average upper
limit for the [Na/Fe] value for FG field stars in the metallicity range covered
by the GCs in our sample (see e.g., Carretta 2013). \label{fig3}}
\end{figure}

\clearpage

\begin{figure}
\centering
\includegraphics[angle=0,scale=.7]{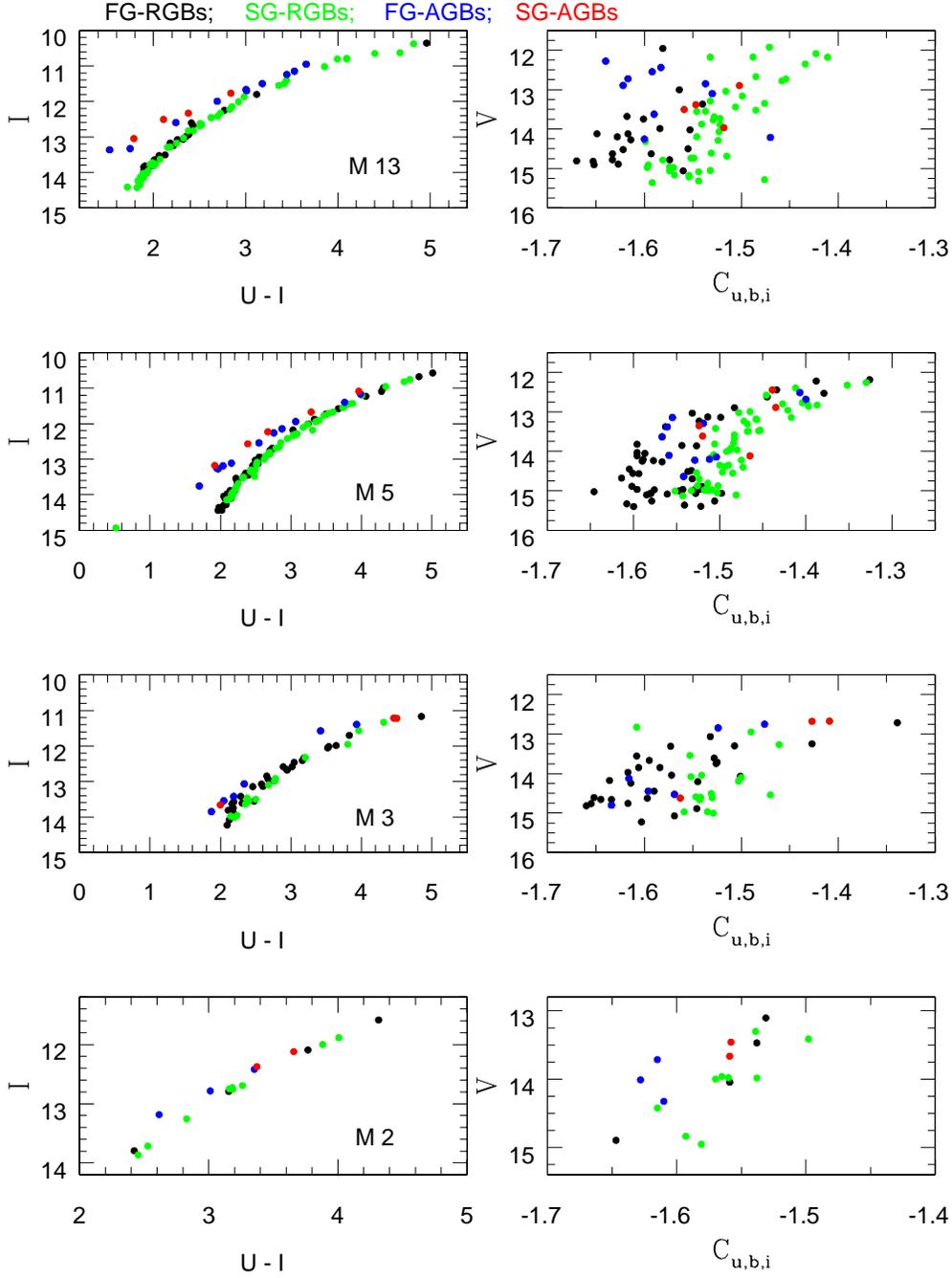}
\caption{Color--magnitude (CMD) diagrams $I$ vs. $(U-I)$ (left panels) and the ($V$,
$C_{u,b,i}$) pseudo-CMDs (right panel) for the metal-poor GCs (from top to
bottom: M 13, M 5, M 3, and M 2) observed by APOGEE. FG- and SG-AGB stars are marked
(blue and red dots, respectively). For comparison, the FG- and SG-RGB stars
(black and green dots, respectively) are also displayed. \label{fig4}}
\end{figure}

\end{document}